\documentclass[fleqn,twoside]{article}
\usepackage{espcrc2}
\usepackage{epsfig}
\usepackage[figuresright]{rotating}

\usepackage{latexsym}
\newcommand{\plus}{\makebox[15pt][c]{$+$}}
\newcommand{\minus}{\makebox[15pt][c]{$-$}}

\newcommand{\err}[2]
{\hskip-0.5em\raisebox{0.08em}{\scriptsize{$\;\begin{array}{@{}l@{}}
\plus\makebox[0.50em][r]{#1\hfill} \\[-0.24em]
\minus\makebox[0.50em][r]{#2\hfill} 
\end{array}$}}}

\newcommand{\AmS}{{\protect\the\textfont2
  A\kern-.1667em\lower.5ex\hbox{M}\kern-.125emS}}
\def\prl#1{Phys.\ Rev.\ Lett.\ {#1}}
\def\prd#1{Phys.\ Rev.\ {D#1}}
\def\plb#1{Phys.\ Lett.\ {B#1}}
\def\npb#1{Nucl.\ Phys.\ {B#1}}
\def\npps#1{Nucl.\ Phys.\ Proc.\ Suppl.\ {#1}}

\def\etal{{\it et al }}

\title{Heavy quark physics from lattice QCD}

\author{Sin\'ead Ryan\address[MCSD]{School of Mathematics,
				    Trinity College,
				    Dublin 2,
				    Ireland}
	}
\begin{document}
\begin{abstract}
I review the current status of lattice calculations of heavy quark quantities.
Particular emphasis is placed on leptonic and semileptonic decay matrix 
elements.
\vspace{1pc}
\end{abstract}

\maketitle

\section{INTRODUCTION}
$B$ and $D$ meson hadronic matrix elements play an important r\^ole in 
determinations of the CKM matrix elements and overconstraining the unitarity
triangle of the Standard Model. Form factors, decay constants and bag
parameters are determined from these hadronic matrix elements which can be 
calculated directly on the lattice. $B$ factories (BaBar and Belle) and
Charm factories (CLEO-c) will reduce experimental uncertainties in 
measurements of decay process, leaving the theoretical calculations as the
dominant uncertainty. Although the focus of lattice heavy quark calculations 
has traditionally been on $B$ physics, new experiments such as CLEO-c 
will determine $f_{D_{(s)}}$ and semileptonic branching 
fractions to a few percent, testing lattice results for these quantities.

In this talk I will report on progress made in the last year in calculations
of heavy quark quantities. Many quantities have now been determined by
different groups using different heavy quark methods. 
The talk is organised as follows. I begin by discussing the status of 
quenched and unquenched calculations of the leptonic decay constants. 
Progress in determinations of the $B$ parameters, $B_{B_{(d,s)}}$, 
which together with $f_{B_{(s)}}$ are relevant to determinations of $|V_{td}|$
and $|V_{ts}|$ and to $B_s$ width and $b$ hadron lifetime measurements, 
is then reviewed. 
Turning to semileptonic decays, I will focus on $B\rightarrow\pi l\nu$ and a 
new method to determine the zero-recoil form factor of 
$B\rightarrow D^\ast l\nu$. A new calculation of the $b$ quark mass and an 
outstanding puzzle in lattice calculations of spin splittings are briefly 
discussed.
\subsection{Heavy quark methods}
The calculations described here employ three different heavy quark methods, 
each of which has different systematic errors. The UKQCD and APE groups use a
nonperturbatively ${\cal O}(a)$-improved Sheikholeslami-Wohlert (SW) action 
(NPSW) at quark masses about that of charm. Results are extrapolated to the 
$b$ quark guided by heavy quark
 effective theory (HQET). A large source of uncertainty comes from this 
extrapolation. It is well known that the nonperturbative 
${\cal O}(a)$-improvement scheme breaks down in the limit
$m_Q\rightarrow\infty$. This was discussed last year by 
Bernard~\cite{bernard00} and an explicit example was given at this conference 
by Kurth and Sommer~\cite{KurthSommer}. They have shown that the 
continuum limit ($a\rightarrow 0$) must be taken before an extrapolation in
heavy mass is attempted.

In nonrelativistic QCD (NRQCD)~\cite{NRQCD} the heavy quark is assumed to be 
nonrelativistic ($(v/c)^2\approx (0.3{\rm GeV}/5.0{\rm GeV})^2$). 
Relativistic momenta are excluded by introducing a finite cut-off such that 
$p\approx m_Qv\ll m_Q$. Then, $|p|/m_Q\ll 1$ and the QCD Lagrangian can be 
expanded in powers of $1/m_Q$. This has been successful for $B$ 
physics, where the quark mass is large, so the expansion is convergent.
The theory is nonrenormalisable but at finite lattice spacing, lattice 
artefacts can be removed by including higher orders in $1/m_Q$ and $a$.

The Fermilab approach (FNAL)~\cite{FnalHQ}, as used in current heavy quark 
calculations, 
is a re-interpretation of the SW action which identifies and 
correctly renormalises nonrelativistic operators present in the SW action. 
Discretisation errors are 
then ${\cal O}(a\Lambda_{QCD})$ and not ${\cal O}(am_Q)$. The existence of a 
continuum limit means a continuum extrapolation is possible. An 
implementation of the full approach, for ${\cal O}(a)$-improvement, has
been done by the Fermilab group~\cite{ZS-fullFNAL}.
\section{LEPTONIC DECAY CONSTANTS}
The $B$ meson decay constant is an input in determinations of the CKM matrix 
element $|V_{td}|$ through $B^0_d-\bar{B}^0_d$ mixing. The mixing, 
$\Delta m_d$, is known quite precisely and the dominant uncertainty comes 
from $f_B\sqrt{B_{B_d}}$.
%
\subsection{The quenched approximation}
\begin{figure}[h]
	\begin{center}
	\epsfxsize=7.5cm\epsfbox{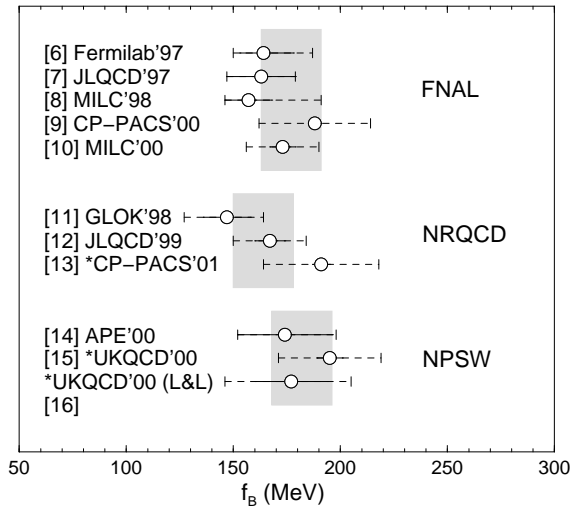}
	\vspace{-1cm}
	\caption{Recent determinations of $f_B$ in the quenched approximation. 
		$\ast$ indicates results updated from Lattice'00. 
		}
	\label{Fig:fBNf0}
	\end{center}
	\vspace{-5ex}
	\end{figure}

\begin{figure}
	\begin{center}
	\epsfxsize=7.5cm\epsfbox{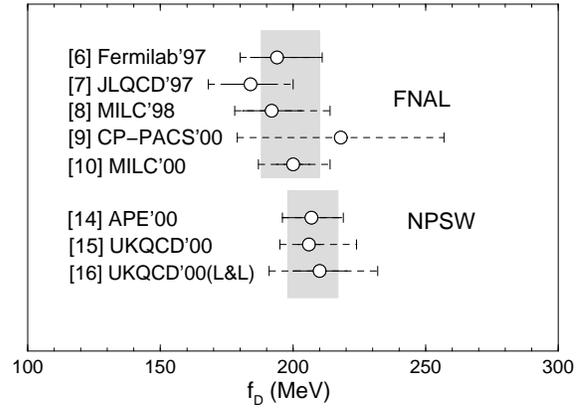}
	\vspace{-1cm}
	\caption{Recent determinations of $f_D$ in the quenched approximation. 
		}
	\label{Fig:fDNf0}
	\end{center}
	\vspace{-5ex}
	\end{figure}

A compilation of recent determinations of $f_B$ and $f_D$ is shown in 
Figs~\ref{Fig:fBNf0} and~\ref{Fig:fDNf0}. 
The calculations are ordered by the heavy quark 
action used. The solid bands indicate my average for the particular heavy 
quark action used and values of $f_B$ obtained for a 
given heavy quark action can be seen to be in good agreement. 
In addition there is a clear 
overlap between results from the different heavy quark actions. 

The most recent calculations of $f_D$ in Fig~\ref{Fig:fDNf0} also show there 
is broad general agreement between
groups and between the two treatments of heavy quarks. 

In 2001 the CP-PACS and UKQCD collaborations both reported updates to 
their quenched determinations of $f_B$, in Refs~\cite{CPPACS:fBNRQCD} and 
~\cite{UKQCD:fBLL,UKQCD:fB} respectively. In this CP-PACS calculation an 
NRQCD action, corrected to ${\cal O}(1/M)$, is used
and $f_B$ is been determined in both the quenched approximation and using 
unquenched gauge configurations. The unquenched result is unchanged and 
is included in Table~\ref{Tab:fBNf2}.
The central value in the quenched approximation is also unchanged from that 
discussed at Lattice'00 by Bernard~\cite{bernard00}. However, the systematic 
error has been revised upwards to reflect the observed large discretisation 
effects. The estimate of the continuum 
value is now $f_B=191\pm 4\pm 27$ MeV and $f_{B_s}= 220\pm 4\pm 31$ MeV, 
where the first error is statistical and
the second is the uncertainty due to discretisation. There are additional
uncertainties in excess of 30\% from setting the lattice scale
and 3\% in $f_{B_s}$ from ambiguities in determining the strange quark mass.

Although the central value is in reasonable agreement with results from other 
groups such large discretisation errors have not been observed previously. 
This scaling analysis is done using fairly coarse lattices, 
$1.017{\rm GeV}\leq a^{-1}_{m_\rho}\leq 1.743{\rm GeV}$ but, nonetheless, 
both GLOK and JLQCD who also use an NRQCD action corrected to 
${\cal O}(1/M)$ and take operator mixing fully into account, as in the CP-PACS 
calculation, see no such scaling violations.
The GLOK group were the first to show that including the 
${\cal O}(\alpha_sa)$ discretisation
correction to the heavy-light axial current leads to a significant decrease 
in scaling violations of $f_B$. Indeed both GLOK and JLQCD find quite mild 
dependence on the lattice spacing, in contrast to CP-PACS.
The calculations differ in the choice of gauge action.
CP-PACS use an Iwasaki RGI action which although it has been shown to have 
small scaling violations, the decay constants seem to be an 
exception~\cite{bernard00,burkhalter00}. 

It is important to understand the reason for the observed scaling violations, 
especially in a quenched calculation, where a motivation for the 
approximation is that it allows us to control and understand other systematic
uncertainties.

The systematic errors in the calculation of decay constants by
Lellouch and Lin in Ref~\cite{UKQCD:fBLL} have been revised 
to include the effects of using different physical quantities to determine 
$am_s$. The changes involved are very small.

The results of the UKQCD Collaboration, reported in 1999 in 
Ref~\cite{UKQCD:fB} and discussed last year in Ref~\cite{bernard00} have been 
substantially revised. The new analysis was described by Maynard at this 
conference~\cite{maynard01}. UKQCD determine $f_B$ using a non-perturbatively 
${\cal O}(a)$-improved action, extrapolated from charm to the bottom quark 
mass. 
The first change is in the current and mass improvement and renormalisation 
coefficients. A consistent set of nonperturbative coefficients, as determined
by Bhattacharya {\it et al}~\cite{fB:BhattCoeffs} is used.
Secondly, the scale is set by the pion decay 
constant rather than $r_0$ as previously. This is a better choice since $r_0$ 
is poorly determined experimentally.
Finally, an extended discussion and 
analysis of the systematic errors and in particular the error in the heavy 
mass extrapolation is included. The mass-dependent normalisations proposed in 
the Fermilab formalism~\cite{FnalHQ} and by Bernard~\cite{bernard00} are 
compared to the nonperturbative normalisation which is used to determine the 
central value. The effect of ${\cal O}(a^2m^2_Q)$ 
lattice artefacts at the charm scale on the extrapolation to the bottom 
quark mass is studied by simultaneously fitting data from two lattice spacings 
to 
\begin{eqnarray}
\Phi (M,a)&=&\gamma\left (1+\frac{\delta}{M}+\frac{\eta}{M^2}\right .
								\nonumber \\
       & & \left .+\epsilon (aM)^2+\xi (aM)^3\right ). \label{Eqn:massextrap}
\end{eqnarray}
This is the usual HQET scaling relation with additional mass-dependent terms.
From the fit a so-called ``quasi-continuum'' result is determined and compared
to fits at individual lattice spacings, the difference being taken as a 
measure of the effect of ${\cal O}(a^2m^2_Q)$ lattice artefacts. While the 
method proposed here is a reasonable one, the size of the error envelope at 
the $B$ meson mass from a fit to Eqn~\ref{Eqn:massextrap} would be a more 
realistic estimate of the uncertainty in this extrapolation. This would  
almost certainly be larger than the error currently quoted by UKQCD. 

I conclude this section with a description of a calculation by Davies 
{\it et al}~\cite{davies01} of $f_B$ at rest and at 
non-zero momentum, using NRQCD. The motivation for such a study comes from 
the need to understand momentum-dependent errors in semileptonic decays, for
which $f_B$ offers an easy place to start. The matrix element 
$\langle 0|A_\mu |B(p)\rangle = f_Bp_\mu$ is studied with temporal, 
$A_0$ and spatial, $A_k$ currents. A number of different 
smearings are investigated and narrow smearing of heavy quarks is found to be 
optimal for moving $B$ mesons. Constrained curve fitting~\cite{guertler01} 
gives reliable results at momenta up to $(pa)^2=16$, in units $(2\pi /L)$ 
and good agreement is found for $f_B$ at zero and 
non-zero momentum. Finally, the discrepancy discussed 
in Ref~\cite{simone96} between $f_B$ from $A_0$ and $A_k$ is resolved by 
including a correct power counting and appropriate normalisation of the 
different $A_\mu$ in the matrix element.
%
\subsection{Unquenching}\label{Sec:fBNf2}
The last year has seen some new calculations of 
decay constants with dynamical quarks. These and other recent calculations are 
summarised in Table~\ref{Tab:fBNf2}. As in 
the quenched case these calculations are performed with a number of heavy 
quark actions. I will discuss the results marked as ``new'' in 
Table~\ref{Tab:fBNf2}. Details of the other calculations can be found in the 
references quoted and in Ref~\cite{bernard00}.
	\begin{table*}[htb]
	\caption{Leptonic decay constants with $N_f=2$. A $\ast$ indicates a new result.}
	\label{Tab:fBNf2}
	\begin{tabular}{@{}|l|cccc|}
	\hline
        Group & $f_B$ (MeV) & $\displaystyle{\frac{f_B^{N_f=2}}{f_B^{N_f=0}}}$ 
        &$f_D$ (MeV) & $\displaystyle{\frac{f_D^{N_f=2}}{f_D^{N_f=0}}}$\\
	\hline
Collins99~\cite{NRQCDfBNf2} &$186(5)(25)(^{+50}_{-0})$  &$\simeq 1.26$ &  &\\
MILC'00$^*$~\cite{MILClat00}    &$191(6)(^{+24}_{-18})(^{+11}_{-0})$   
		       &$\simeq 1.10$ 
                       &$215(5)(^{+17}_{-13})(^{+8}_{-0})$ 
		       &$\simeq 1.08$ \\
MILC'01$^*$ ($N_f=2+1$)~\cite{bernard01}    
		       &  & $1.23(3)(11)$  &   &\\
CP-PACS'00(FNAL)~\cite{CPPACS:fBKKM}     
		       &$208(10)(29)$ & 1.11(6) & 225(14)(40) & 1.03(6)\\
CP-PACS'00(NR)~\cite{CPPACS:fBNRQCD} 
		       & $204(8)(29)(^{+44}_{-0})$ 
		       & $1.10(5)$ &  &\\     
JLQCD$^*$~\cite{yamada-talk}  
		       & $190(14)(7)$  &$\simeq 1.14$ &  &\\
	\hline 
	\end{tabular}\vskip -.15truein
	\end{table*}

Yamada~\cite{yamada-talk}, for the JLQCD Collaboration, presented 
preliminary results from a calculation of $f_{B_{(s)}}$ using NRQCD, 
including all $1/M$ corrections. A nonperturbatively ${\cal O}(a)$-improved 
Wilson action for the sea and valence light quarks and a plaquette gauge 
action are used. They compare values of $f_B$ in the quenched approximation 
using a perturbative (tadpole-improved) and nonperturbative value of $c_{SW}$.
 The difference is found to be negligible. Using nonperturbative improvement
coefficients, the quenched value of $f_B$ is compared to the two flavour
result ($N_f=2$). $f_B$ is larger in the unquenched theory but the size of 
the difference depends on the
type of valence chiral extrapolation used. Linear, quadratic and $\log$ fits 
were compared and the latter two yield good $\chi^2/N_{\rm df}$. 
The $\log$ fits are motivated by chiral perturbation theory~\cite{sharpe-zhang}
and to my knowledge this fit form has not been explored by other groups, for
this particular quantity. The results from JLQCD indicate a careful analysis
of the chiral behaviour is warranted. For their preliminary result, in 
Table~\ref{Tab:fBNf2} the central value is an average of the quadratic and 
$\log$ fits and the second error is the difference from the linear 
extrapolation. They are gathering more statistics for a
detailed study of the chiral extrapolation. 
The group also presented results for the $B$ parameters which I will discuss 
below. 

The ``MILC'00'' result was discussed in some detail in Ref~\cite{bernard00}. 
I refer the reader there for more details and turn to the newest result from 
MILC. Preliminary results for $f_B$ and ratios were presented at this 
conference by Bernard~\cite{bernard01}. An improved Kogut-Susskind action 
(``Asqtad'') with $N_f=2+1$ was used. 
For heavy quarks they use a tadpole-improved SW action with the Fermilab
nonrelativistic interpretation. The dependence of $f_B$ on 
dynamical $(m_\pi /m_\rho )^2$ is investigated with constant and linear fits.
The valence chiral extrapolation is seen to be well controlled with the 
difference between linear and quadratic fits being quite small, 
in contrast to the JLQCD result. 
Since both calculations are preliminary, definite 
conclusions are premature and further investigation is required. 
The 
currents have not been normalised and so I do not include the 
value of $f_B$ in Table~\ref{Tab:fBNf2}. In a ratio of decay constants the 
normalisation factors largely cancel, as do many systematic uncertainties. MILC
find a mild dependence on both $N_f$ and $a^{-1}$ in the ratio $f_{B_s}/f_B$.

Unquenched simulations (with $N_f=2$) of $D$ meson decay constants have been 
done by CP-PACS and MILC as reported in Table~\ref{Tab:fBNf2}. Both groups 
use the Fermilab nonrelativistic interpretation for the heavy quarks and 
an improved Wilson action for
light quarks, although with different gauge actions. The results agree 
within the quoted uncertainties. The results at three lattice spacings do not
show scaling and CP-PACS note that $f_D$ may in fact be smaller
than their reported value due to discretisation effects. They are able to 
conclude however that the dynamical effect for $D$ mesons is smaller than for 
$B$ mesons. This is supported by the MILC data. 

It is a difficult task to combine results given the very different systematic 
errors in different calculations. Nonetheless, I 
believe the estimates below represent a summary of the current status.
\begin{tabbing}
Quenched   \hspace{2cm}     \= Unquenched \\
$f_B = 173(23)$ MeV     \> $f_B      =  198(30)$ MeV \\
$f_{B_s}$ = 200(20) MeV \> $f_{B_s}$ = 230(30) MeV \\
$f_D = 203(14)$ MeV     \> $f_D      = 226(15) $  MeV \\   
$f_{D_s} = 230(14)$ MeV \> $f_{D_s}  = 250(30)$ MeV \\
$f_{B_s}/f_B = 1.15(3)$ \> $f_{B_s}/f_B =  1.16(5)$ \\ 
$f_{D_s}/f_D = 1.12(2)$ \> $f_{D_s}/f_D = 1.12(4)$ 
\end{tabbing}
Of these, only $f_{D_s}$ has been measured experimentally. A recent average 
is $f_{D_s}=280\pm 48$ MeV~\cite{PDG2000}, in agreement with the average 
above. 
\section{ $B$ PARAMETERS }
In comparison to $f_B$ lattice calculations of $B$ parameters have not 
received as much attention. There is however, progress to report this year.

A new calculation of $\Delta B=0$ operators by APE is described in 
Ref~\cite{apedb0}. These matrix elements are important in lifetime ratios of 
$B$ mesons.
Reyes~\cite{reyes01} presented details of a quenched calculation by the APE 
collaboration of the matrix elements of $\Delta B=2$ operators. A more 
detailed description has recently appeared in Ref~\cite{APE:Bparams}.
Results from a relativistic simulation (at the charm quark mass) are combined
with results in the static limit and interpolated to the $b$ quark mass.
The operators in the relativistic and static theories are matched at NLO, 
including the anomalous dimension matrix up to two loops and the one-loop
matching for the $\Delta B=2$ operators. This is the first time
 such a consistent matching has been done and since the $\Delta B=2$ 
operators are determined by interpolation rather than extrapolation the 
systematic error is significantly reduced. 
The result is $B_{B_d}(m_b)=0.87(4)(3)(0)\left (\err{3}{4}\right )$, 
where the errors are statistical, 
systematic errors in the static calculation, error in the renormalisation in 
QCD and a combined error due to the uncertainty in $a^{-1}$ and in the 
improvement coefficient of the axial current. Not included here is a full
estimate of the discretisation error although it is discussed in the text. 
The operators are not ${\cal O}(a)$-improved and the static and SW
calculations are done at different lattice spacings. 
The discretisation error due to the unimproved current is included in the 
third error bar, any remaining $a$-dependence is difficult to 
estimate with just one lattice spacing.

JLQCD~\cite{yamada-talk} use NRQCD determine $B$ parameters from the quenched 
and unquenched data sets, already described in Section~\ref{Sec:fBNf2}. 
They find a negligible difference in the two results. 
The matrix elements are renormalised at the one-loop level in perturbation 
theory. 
The light quark mass dependence is mild and no significant dependence on the
choice of fit function is observed. The dependence on the
quantity used to set the scale and the determination of the strange quark
mass are included in the error budget. Their result, with $N_f=2$ is
$B_{B_d}(m_b)=0.872(39)(19)(4)$. The errors are statistical, the difference
from four methods to determine $B_B$ and the chiral extrapolation.
Power counting leads to an additional 8\%
systematic error. Sharpe has estimated~\cite{SharpeQCPTh} the effect of the 
quenched approximation, from quenched chiral perturbation theory, to be 
$\approx 10\%$. This calculation
suggests that this 10\% may be an upper bound on the effect of unquenching 
but as
this is a preliminary result I think it is best to leave the 10\% uncertainty
and await the final result and subsequent confirmation by other groups.

Taking these two new (albeit preliminary) results into account I estimate the
renormalisation group invariant $B$ parameters to be
\begin{eqnarray*}
\hat{B}_{B_d}	  &=& 1.30(12)(< 10\%)\\
\hat{B}_{B_s}	  &=& 1.34(10)(<10\%) \\
B_{B_s}/B_{B_d} &=& 1.03(3)	\\
\xi = f_{B_s}\sqrt{B_{B_s}}/f_B\sqrt{B_{B_d}} &=& 1.15(6)
						\left (\err{7}{0}\right )
\end{eqnarray*}
The second, asymmetric error on $\xi$ takes into account the preliminary
result from JLQCD. This suggests that in the unquenched calculation of $f_B$
chiral logs are important in the chiral extrapolation. 
$f_{B_s}$ is uneffected so the value of $\xi$ in future may be 
 larger than presently quoted. 
\section{SEMILEPTONIC DECAYS}
A determination of $|V_{ub}|$ and $|V_{cb}|$ can be made by combining
experimental measurements of exclusive branching fractions of 
$B\rightarrow\pi ,\rho l\nu$ and $B\rightarrow D^{(\ast )}l\nu$ with
lattice calculations of the hadronic matrix elements (or corresponding form 
factors). In practice, lattice calculations of these matrix elements, with 
experimentally favoured kinematics present various problems. I will report
on some recent progress in calculations of 
$B(D)\rightarrow\pi ,\rho l\nu$ and $B\rightarrow D^{(\ast )}l\nu$.
Preliminary results, from UKQCD for the decay
$B\rightarrow\rho l\nu$ were presented at this conference by
Gill~\cite{UKQCDB2rho}.
	\section{$B\rightarrow\pi l\nu$}
The hadronic matrix elements parameterising the decay of heavy-light to light 
mesons are not protected by flavour symmetries and hence $|V_{ub}|$ is 
dominated by theoretical uncertainties. 

$|V_{ub}|$ is determined from the decay rate 
\begin{equation}
\frac{d\Gamma}{dp} = \frac{G_F^2|V_{ub}|^2}{24\pi^3}\frac{2m_Bp^4|f_+(E)|^2}
						 {E}\label{Eqn:B2pirate},
\end{equation}
where the form factors, $f_+(E)$ and $f_0(E)$ parameterise the matrix 
elements in the standard way
\begin{eqnarray}
\langle\pi (p_\pi )|{\cal V}^\mu |B(p_B)\rangle &=& 
				f_+(E)\left [p_B+p_\pi\right .\nonumber\\
-\left . \frac{m_B^2-m_\pi^2}{q^2}q\right ]^\mu 
	 &+&f_0(E)\frac{m_B^2-m_\pi^2}{q^2}q^\mu    \label{Eqn:tradFF}
\end{eqnarray}
I will focus on four recent calculations by UKQCD~\cite{UKQCDB2pi}, 
APE~\cite{APEB2pi}, FNAL~\cite{FNALB2pi} and JLQCD~\cite{JLQCDB2pi}. UKQCD 
and APE extrapolate from quark masses around charm to the bottom quark mass. 
The Fermilab group use the Fermilab formalism for heavy quarks and JLQCD 
use NRQCD. Thus it 
is possible to investigate the agreement of different analyses and treatments
of heavy quarks. Table~\ref{Tab:B2pi} lists some details of these 
analyses. A number of improvements, compared to previous calculations are 
included. All groups make the chiral extrapolation and the renormalisation
of matrix elements is done nonperturbatively or at one loop in perturbation
theory. The lattice spacing dependence is studied by the Fermilab group who
also simulate directly at the $b$ quark mass so there is no uncertainty due 
to extrapolation of heavy quark mass in the calculation. This is, of course,
also true of the JLQCD calculation.
%
	\begin{table*}[htb]
	\begin{center}
	\caption{Overview of the $B\rightarrow\pi l\nu$ analyses. PT is 
perturbation theory and NP is nonperturbative.}
	\label{Tab:B2pi}
	\begin{tabular}{@{}|l|llll|}
	\hline  
		& APE	 	 & UKQCD 	  & FNAL 	     &JLQCD\\
	\hline
simulate at $m_b$ & no		 & no     	  & yes 	     &yes \\
quark mass & ${\rm M}_{\rm rest}$ & ${\rm M}_{\rm rest}$ & 
		     ${\rm M}_{\rm kinetic}$ & ${\rm M}_{\rm kinetic}$\\
chiral extrap.  & yes   & yes    & yes     & yes \\
lattice spacings    & 1     	 &   1     	  & 3 	     & 1  \\
matching        & NP at $m_q=0$  & NP at $m_q=0$  
                         & mostly NP:$\sqrt{Z_{V^{uu}}Z_{V^{bb}}}$  & PT \\
    &includes KLM term & includes KLM term 
                            &PT: radiative corrections   & 1-loop  \\
	&	&	&at 1-loop & \\
	\hline
	\end{tabular}
	\end{center}\vskip -.15truein
	\end{table*}

\begin{figure}[h]
	\begin{center}
	\epsfxsize=7.5cm\epsfbox{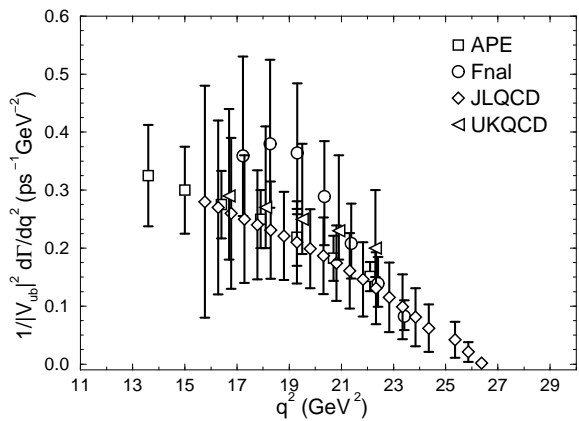}
	\vspace{-1cm}
	\caption{Comparison of differential decay rates.}
	\label{Fig:B2piRate}\vspace{-0.05truein}
	\end{center}
	\vspace{-3ex}
	\end{figure}

Fig~\ref{Fig:B2piRate} shows the differential decay rate as a function of the
recoil momentum, $q^2=m_B^2-2m_BE+m_\pi^2$, from the four groups listed above.
 The error bars are
the statistical and systematic errors in each calculation. The data show 
general agreement between the four groups, within the quoted uncertainties. 
The major shortcoming of all lattice calculations can also be seen from this
plot: the bulk of the experimental data is at $q^2$ close to zero. To avoid
a model-dependent extrapolation in $q^2$ new techniques are needed to increase
the range of momenta, at currently accessible lattice spacings. Alternatively
new experiments may be able to extend the current kinematic range for this 
decay.
	\vspace{-2ex}
\subsection{Fermilab and JLQCD}
Rather than use the parameterisation of Eqn~\ref{Eqn:tradFF} both
Fermilab and JLQCD use forms motivated by HQET.
\begin{eqnarray*}
\langle\pi (p_\pi )|{\cal V}^\mu |B(p_B)\rangle &=& 
    \sqrt{2m_B}\left [v^\mu f_\parallel (E) \right .\\
    & & \left .+ p^\mu_\perp f_\perp(E)\right ]\\
&=& 2\left [ f_1(E)v^\mu +f_2(E)\frac{p_\pi^\mu}{E}\right ] ,
\end{eqnarray*}
where $v=p_B/m_B$ and $p_\perp =p_\pi -Ev$.
The first parameterisation is used by Fermilab and the second by JLQCD.
The traditional form factors, $f_+$ and $f_0$ are easily obtained from
$f_\parallel$ and $f_\perp$ or $f_1$ and $f_2$. These are useful quantities to
focus on. Considering the $m_\pi ,E\rightarrow 0$ limit yields 
\begin{equation}
f_\parallel = \frac{f_B\sqrt{m_B}}{\sqrt{2}f_\pi}  \label{Eqn:SPT}
\end{equation}
\begin{equation}
f_\perp = \frac{f_B^\ast\sqrt{m_B^\ast}}{\sqrt{2}f_\pi}g_{BB^\ast\pi}
\frac{2m_B}{m_{B^\ast}^2 -q^2} 
\end{equation}
through ${\cal O}(1/m_Q)$ in the heavy-quark expansion. Similar expressions can
be written down for $f_{(1,2)}$.
In addition, these form factors have simple descriptions in HQET and so are
natural in the Fermilab and NRQCD approaches. Finally, they emerge directly 
from the lattice calculation and can be analysed directly, forming $f_{(+,0)}$
at the end. 

\begin{figure}[h]
	\begin{center}
	\epsfxsize=7.5cm\epsfbox{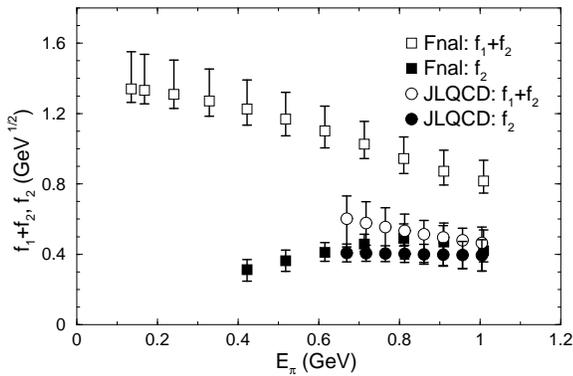}
	\vspace{-1cm}
	\caption{Comparison of JLQCD and Fermilab form factors}
	\label{Fig:f1f2_vsE}\vspace{-0.05truein}
	\end{center}
	\vspace{-5ex}
	\end{figure}

In Ref~\cite{JLQCDB2pi} a discrepancy between one of the Fermilab and
 JLQCD form factors was discussed. Fig~\ref{Fig:f1f2_vsE} shows this 
discrepancy in $f_1+f_2$. This form factor yields $f_0$ which does 
not contribute to the rate. However, one would expect the NRQCD and Fermilab
approaches to be in agreement especially as the simulation parameters are 
also similar. 
In fact the functional form 
of the chiral extrapolation and the range of light quark masses has a 
significant effect on the final values. JLQCD use a linear extrapolation with 
a range of light quark masses $0.034\leq am_0\leq 0.072$ and Fermilab 
determine the central value with from a quadratic fit with light quarks in the 
range $0.024\leq am_0\leq 0.059$. With this dependence on the light quark 
masses and the functional form of the chiral extrapolation, the 
$\displaystyle{\left (\err{14}{22}\right)\%}$ uncertainty due to chiral
extrapolation assigned to $d\Gamma /dp$ in the Fermilab analysis seems 
prudent. It is in fact the largest systematic error in that analysis.
Lighter quark masses and higher statistics would certainly help to reduce this
uncertainty.
\subsection{Heavy mass dependence}
\begin{figure}[h]
	\begin{center}
	\epsfxsize=7.5cm\epsfbox{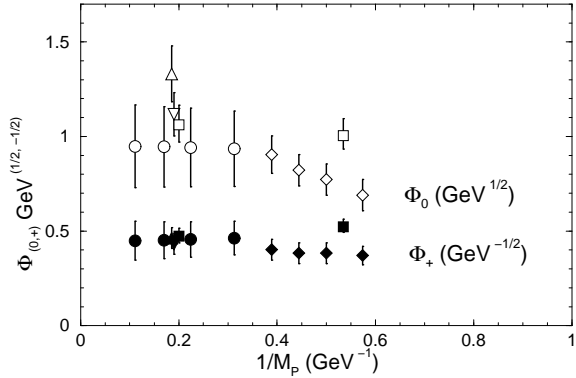}
	\vspace{-1cm}
	\caption{Mass dependence of JLQCD ($\circ$), APE ($\diamond$) and 
			Fermilab ($\Box$) form factors.}
	\label{Fig:B2piMassdep}
	\end{center}\vskip -.15truein
	\end{figure}

Turning now to the heavy mass dependence. This has been studied by JLQCD and 
APE although not in the same region of heavy quark mass. JLQCD simulate at 
four heavy quark masses around the $b$ quark. Plotting the quantities
$\Phi_+\equiv (\alpha_s(M_P)/\alpha_s(M_B))^{-2/11}f_+/\sqrt{M_P}$ and
$\Phi_0 \equiv (\alpha_s(M_P)/\alpha_s(M_B))^{-2/11}f_0\sqrt{M_P}$ at fixed 
$E$ they see no dependence on the heavy mass. This analysis is restricted to 
the bottom quark regime.
On the other hand, APE find a negative slope when plotting the same quantities
versus $1/M_P$ with heavy quarks at around the charm mass. These differences 
in slope can be seen in Figure~\ref{Fig:B2piMassdep}.
The 
Fermilab nonrelativistic approach is valid at arbitrary quark mass. 
Fig~\ref{Fig:B2piMassdep}
compares the JLQCD, APE and Fermilab data where only Fermilab have 
simulated directly at both the charm and bottom quark masses. 
The figure reveals that the Fermilab data agree with JLQCD at $M_B$ but show
slope as $1/M_P\rightarrow 1/M_D$. The slope of the mass dependence is 
gentler than that observed by APE agreeing more with a linear extrapolation
of the APE data to $M_B$.

I make some final remarks in this section about the soft pion theorem (SPT). 
The relation applies in the limit $m_\pi ,E\rightarrow 0$ as given by 
Eqn~\ref{Eqn:SPT}.
The first remark is that the limit in which this relation holds implies $f_0$
should be evaluated at $m_B^2$ rather than $q^2_{\rm max}=m_B^2-m_\pi^2$.

The question of whether or not lattice calculations satisfy this relation has
been discussed in previous reviews~\cite{bernard00,onogi97}. In the analyses
discussed here both APE and UKQCD report that the relation in Eqn~\ref{Eqn:SPT}
holds. In the JLQCD and Fermilab results this relation does not hold. It is 
not at all clear what the reasons for this are. 
The APE group use the soft pion theorem as a criterion for
choosing the functional form (quadratic rather than linear) of the heavy mass 
extrapolation so it is perhaps not surprising then that the relation holds. 
UKQCD also choose a quadratic form for the heavy mass extrapolation and there
is a dependence on the pole form introduced when interpolation to a 
common set of $q^2$. JLQCD observe that a $\sqrt{m_q}$ term in the chiral 
extrapolation raises the value of the soft pion limit. Thus the problem is
dependent on the form of the extrapolation and the interpolation to 
fixed $E$. 
In addition, Onogi has pointed out~\cite{onogi97} that the perturbative 
uncertainty in the renormalisation coefficients of the heavy-light currents
may be contributing to the SPT discrepancy. This year Onogi~\cite{onogi01} 
reported on a nonperturbative determination of the ratio, $Z_A/Z_V$ with
static heavy quarks and SW light quarks. The result brings the two sides
of Eqn~\ref{Eqn:SPT} into better agreement, but does not explain completely 
the difference.

Finally, the Fermilab group note that while the lattice data for $f_0$ do not 
satisfy Eqn~\ref{Eqn:SPT} the experimental rate at this point
goes to zero and the lattice systematics (mostly due to chiral extrapolation) 
increase at the soft pion limit. For these reasons cuts are imposed on the 
range of energy considered such that $E\geq 0.424$GeV. Although the 
phenomenologically interesting region is far from $q^2_{max}$ the reasons for 
the violations of the SPT are important. It is a good indication of the control
of systematic errors. 

The sources and severity of the different systematic errors in the
calculation of differential decay rates and their
contribution to the theoretical uncertainty in $|V_{ub}|$ were tabulated in
the Fermilab analysis. The details are in Table~\ref{Tab:B2pierrors}. 
	\vspace{-2ex}
        \begin{table}[htb]
        \begin{center}
        \caption{Sources of systematic error, in percent, in
                 $B(D)\rightarrow\pi l\nu$ from the Fermilab
                 analysis~\cite{FNALB2pi}.}
        \label{Tab:B2pierrors}
        \begin{tabular}{@{}|l|cccc|}
        \hline
Error              & $T_B$  & $|V_{ub}|$  & $T_D$ &  $|V_{cd}|$ \\
\hline
		   &		  &		&	& \\
statistical        &$^{+27}_{-9}$ &$^{+14}_{-5}$ &$^{+17}_{-8}$
                                                          &$^{+9}_{-4}$\\
excited state      &$\pm 6$    &$\pm 3$  &$\pm 6$         &$\pm 3$  \\
$\vec{p}$ extrapol$^n$ & $\pm 10$    &$\pm 5$  &$\pm 9$  &$\pm 5$\\
$m_q$ extrapol$^n$ & $^{+16}_{-22}$  &$^{+8}_{-11}$ &$^{+3}_{-18}$ &
                                                          $^{+2}_{-9}$ \\
adjusting $m_Q$    & $\pm 6$    &$\pm 3$  &$\pm 2$  &$\pm 1$    \\
HQET matching      & $\pm 10$    &$\pm 5$  &$\pm 10$  &$\pm 5$  \\
$a$ dependence     & $^{+16}_{-3}$   &$^{+8}_{-2}$ &$^{+23}_{-6}$
                                                          &$^{+11}_{-3}$\\
definition of $a$  &$\pm 11$    &$\pm 6$  &$\pm 4$  &$\pm 2$    \\
\hline
Total systematic   & 30 & 15 & $^{+28}_{-24}$ & $^{+14}_{-12}$ \\
                   &    &       &       &       \\
Total (stat$\otimes$sys) & $^{+40}_{-31}$ & $^{+20}_{-16}$ & $^{+32}_{-26}$
                                         &$^{+16}_{-13}$\\
		   &		  &		&	& \\
\hline
\end{tabular}
\end{center}
\vspace{-6ex}
\end{table}
$T_B$
is defined from Eqn~\ref{Eqn:B2pirate} as $T_B=2m_Bp^4|f_+(E)|^2/E$ and
similarly for $T_D$. The $D\rightarrow\pi l\nu$ calculation is discussed in 
Section~\ref{Sec:D2pi}.
	\section{$D\rightarrow\pi (K)l\nu$}\label{Sec:D2pi}
This decay has traditionally received less attention than the 
$B\rightarrow\pi l\nu$. With the planned charm factories at CLEO-c and 
elsewhere it is increasingly important to focus on $D$ meson decays. This is
a region of quark mass where Fermilab's nonrelativistic interpretation 
and the ${\cal O}(a)$-improved approach should be in agreement. 
A nice advantage in this decay is that lattice calculations can 
reach the entire kinematical range with no extrapolation in $q^2$. This 
removes what is one of the major problems with the $B$ meson decay.

Both APE and Fermilab have reported results for the decay 
$D\rightarrow\pi l\nu$. In addition the APE group presented their results for
$D\rightarrow Kl\nu$. Preliminary results for this transition were presented
by the Fermilab group in Ref~\cite{FnalD2K}. 

As already indicated by Fig~\ref{Fig:B2piMassdep} the results of the Fermilab
and APE analysis do not agree at the $D$ meson. 
The difference is even greater when one looks 
at the differential decay rate since $(d\Gamma /dq^2)\propto |f_+(E)|^2$.
There are a number of differences in both the action and 
improvement coefficients and the analysis of the groups. The Fermilab group
define the quark mass from the kinetic mass rather than the rest mass, as 
APE does. In lattice calculations of heavy quark systems the kinetic and rest 
masses do not agree. The kinetic mass describes the dynamics of a 
nonrelativistic system and this choice is advocated in the Fermilab formalism
 (as well, of course, as in NRQCD). However, at masses around that of charm 
the difference between rest and kinetic mass is not as great and the choice of
one definition rather than the other cannot, I believe, explain the lack of 
agreement. Further differences arise in the current renormalistation. In the
Fermilab scheme the bulk of this is done nonperturbatively in a fully 
mass-dependent way. The remaining matching from lattice HQET to continuum HQET
is done at one-loop level in perturbation theory. APE use a nonperturbative 
determination of the matching coefficients for massless quarks with the 
so-called KLM term to correct for ${\cal O}(ma)$ effects. In addition there 
are differences such as the use of a quadratic (Fermilab) and a linear (APE) 
chiral extrapolation. It may be that these differences combine to produce
the large discrepancy seen in the form factors. With 
experiments soon to test lattice predictions it is important to understand 
what is going on. I would also like to note that the UKQCD collaboration 
do have preliminary results. Since they are not final I have not included 
them in this discussion although interestingly they lie between the Fermilab 
and APE results. 
	\section{$B\rightarrow D^{(\ast )}l\nu$}
New results were presented in two talks at this conference. 
Simone~\cite{simone:lat01} reported on work by the Fermilab group to 
measure the form factors for this decay at zero recoil. 
Lacagnina~\cite{lacagnina:lat01} reported preliminary results from 
UKQCD for the shape of the Isgur-Wise function determined 
from $B\rightarrow D^\ast l\nu$ and $B\rightarrow Dl\nu$. I will focus the 
Fermilab results since they were (almost) final at the time of the 
conference and the calculation has recently been completed and appeared in 
Ref~\cite{B2Dstar}. 

At the zero recoil point the simple relation
${\cal F}_{B\rightarrow D^\ast}(1) = h_{A_1}(1)$
holds and heavy quark symmetry constrains $h_{A_1}$ to have the heavy quark 
expansion
\begin{equation}
h_{A_1} = \eta_A\left [ 1-\frac{l_V}{(2m_c)^2}+\frac{2l_A}{2m_c2m_b}-
	\frac{l_P}{(2m_b)^2}\right ]. \label{Eqn:hA1}
\end{equation}
In Eqn~\ref{Eqn:hA1} the radiative correction, $\eta_A$ is known to two-loop
level from a calculation by Czarnecki and Melnikov~\cite{B2dstar:Czarnecki}.
The three $l$s are hadronic matrix elements of the HQET and are calculable in 
lattice QCD. 

To determine these matrix elements, for the first time in a lattice 
calculation, a new method is introduced. It exploits an idea developed in a 
previous paper where double ratios of matrix elements were used to determine
the form factors of $B\rightarrow Dl\nu$ at zero recoil~\cite{Fnal:B2D}. 
Three such double ratios are employed to determine
$l_V,l_P$ and $l_A$.
\begin{eqnarray*}
{\cal R}_+&=&\hspace{-.1cm}\frac{\langle D|\bar{c}\gamma^4b|\bar{B}\rangle
	            \langle\bar{B}|\bar{b}\gamma^4c|D\rangle}{
   \langle D|\bar{c}\gamma^4c|D\rangle\langle\bar{B}|\bar{b}\gamma^4b|B\rangle}
	  = |h_+(1)|^2 \\
{\cal R}_1 &=&\hspace{-.1cm} \frac{\langle D^\ast|\bar{c}\gamma^4b|\bar{B}^\ast\rangle
		\langle\bar{B}^\ast|\bar{b}\gamma^4c|D^\ast\rangle}{ 
   	        \langle D^\ast|\bar{c}\gamma^4c|D^\ast\rangle
		      \langle\bar{B}^\ast|\bar{b}\gamma^4b|B^\ast\rangle}
	  =  |h_1(1)|^2 \\
{\cal R}_{A_1} &=&\hspace{-.1cm} \frac{\langle D^\ast|\bar{c}\gamma_j\gamma_5b|\bar{B}\rangle
	\langle\bar{B}^\ast|\bar{b}\gamma_j\gamma_5c|D\rangle}{ 
	\langle D^\ast|\bar{c}\gamma_j\gamma_5c|D\rangle
		\langle\bar{B}^\ast|\bar{b}\gamma_j\gamma_5b|B\rangle} 
	= |\check{h}_{A_1}(1)|^2
\end{eqnarray*}
where $|\check{h}_{A_1}(1)|^2$ is defined by 
$\frac{h_{A_1}^{\bar{B}\rightarrow D^\ast}(1)
		     h_{A_1}^{D\rightarrow\bar{B}^\ast}(1) }{
		h_{A_1}^{D\rightarrow D^\ast}(1)
	        h_{A_1}^{\bar{B}\rightarrow \bar{B}^\ast}(1) }$.
The matrix elements $l_P,l_V$ and $l_A$ are obtained from the heavy quark 
expansions of $h_+(1),h_1(1)$ and $\check{h}_{A_1}(1)$ respectively. The 
required form factor, $h_{A_1}(1)$ can then be reconstructed using 
Eqn~\ref{Eqn:hA1}. The Fermilab group report a final result 
\begin{equation}
{\cal F}_{B\rightarrow D^\ast l\nu}(1) = 0.9130^{+0.0238}_{-0.0173}{}^{ +0.0171}_{ -0.0302}
\end{equation}
where the first error is statistical and fitting and the second systematic, 
resulting from matching lattice gauge theory and HQET to QCD, lattice spacing
dependence, light quark mass dependence and the quenched approximation. 

The bulk of the matching cancels in the double ratios and the remaining 
short-distance coefficients that match lattice gauge theory to QCD and HQET to 
QCD are small and are calculated at one-loop level in perturbation theory. Lattice spacing
dependence is studied using three lattices in the range $5.7\leq\beta\leq 6.1$.
$h_{A_1}(1)$ is extrapolated linearly in $m_\pi^2$ to the chiral limit, showing
a downward trend. The quoted uncertainty is a result of including nonlinear 
terms present in the chiral expansion. The chiral extrapolation is in 
fact the largest source of uncertainty. Finally the error due to quenching is
estimated by summing the effect on the running coupling, $\alpha_s$ and
allowing an additional $10\%$ uncertainty. 

This is a precision calculation of a phenomenologically interesting
quantity. The result agrees well with those from other methods 
(non-relativistic quark models~\cite{b2Dstar:neubert} and a zero-recoil sum 
rule~\cite{B2Dstar:BSU,B2Dstar:SUV}). Combining the result with 
measurements by CLEO~\cite{CLEO:B2Dstar}, LEP~\cite{LEP:B2Dstar} and 
Belle~\cite{Belle:B2Dstar} implies
$$
10^3|V_{cb}| = \left\{ \begin{array}{cc}
			45.9\pm 2.4\err{1.8}{1.4}\\
			38.7\pm 1.8\err{1.5}{1.2}\\
			39.3\pm 2.5\err{1.6}{1.3}\end{array}\right .
$$
%
\section{$m_b$}
\begin{figure}[h]
        \begin{center}
        \epsfxsize=7.5cm\epsfbox{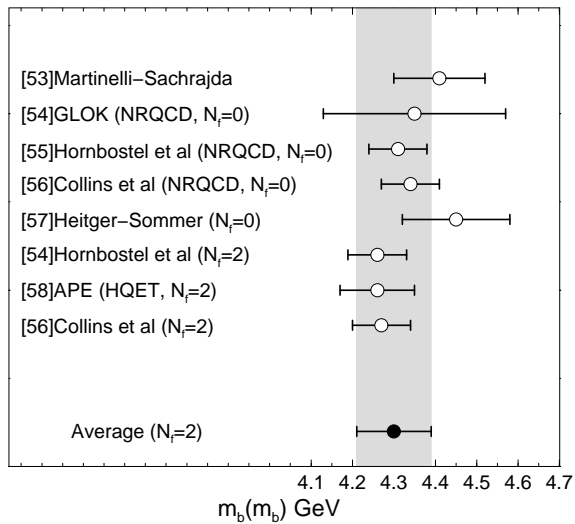}
        \vspace{-1cm}
        \caption{Lattice calculations of $m_b$.}
        \label{Fig:mb}\vspace{-0.05truein}
        \end{center}\vskip -.15truein
        \end{figure}

A new method to determine the $b$ quark mass with nonperturbative accuracy was
presented by Sommer~\cite{Sommer01}. This method has the advantage of 
avoiding the perturbative subtraction of power law divergences in the relation
$\hat{m}_b(\mu )=Z_{\rm cont}(\mu )M_{\rm pole}$ and therefore, taking the 
continuum limit is 
possible with this method. Fig~\ref{Fig:mb} summarises the current status,
including quenched and unquenched calculations and Sommer's preliminary result.
This is in good agreement with previous 
determinations. I estimate an average of these calculations to be 
$\overline{m}_b(\overline{m}_b) = 4.30(10)\mbox{ GeV}$.
\section{SPECTROSCOPY}
In this section I will focus on one quantity: the hyperfine splitting in both
heavy-light and quarkonia systems. Lattice calculations of spectroscopic 
quantities are generally better controlled than matrix element calculations of
similar scope. A major advantage is that no renormalisation is required. 
However, the hyperfine splitting (HFS) remains an exception to this rule. 

In the heavy-light sector it is experimentally observed that the
value of the HFS remains constant for all flavours. In practice, in the
meson sector, lattice results are suppressed relative to experiment by
as much as 20\%(40\%) for $D$($B$) mesons. At this conference 
Tsutsui~\cite{tsutsui01} described a calculation of heavy quark expansion
parameters using NRQCD which includes results for the heavy-light meson and 
baryon splittings. The meson splitting is $\simeq 30\%$ below the 
experimental value as in other calculations while this discrepancy
does not exist in the baryon splittings.

It is argued that the discrepancy is a 
quenching effect since the meson splitting is proportional to the wavefunction
at the orgin which is suppressed in the quenched approximation and
to the strong coupling which runs differently in the full and quenched 
theories. This however, has not been verified by an unquenched calculation
of these splittings.
Lewis~\cite{HFS:Lewis} presented results for 
the charmed baryon spectrum (both $QQq$ and 
$Qqq$) using anisotropic and NRQCD actions. Despite larger statistical errors 
than in the meson spectrum they see no suppression of the HFS. In fact there
may be evidence that it is overestimated although better data are required to 
make a conclusive statement.

Turning to the charmonium and bottomonium spectra. Garcia-Perez reported 
preliminary results on behalf of the QCDTARO collaboration~\cite{HFS:TARO}. 
They have calculated the HFS on a very fine lattice, $\beta =6.6$, using the 
SW action. 
They take the continuum limit using UKQCD data at $\beta =6.0,6.2$ and 
observe significant lattice spacing dependence. 
They are currently producing their own configurations at a range of $\beta$
values to make a more consistent continuum extrapolation. 
New results for the bottomonium spectrum and HFS were presented by 
Manke~\cite{HFS:Columbia}. The calculation is 
done with an anisotropic relativistic heavy quark action. Two levels of 
anisotropy are used in the quenched approximation. They also find significant 
scaling violations. 

\section{CONCLUSIONS}
Lattice calculations offer the prospect of model-independent determinations 
of hadronic matrix elements. The effect of unquenching in calculations of 
leptonic decay constants and their ratios is becoming more precise. 

For the semileptonic decay $B\rightarrow\pi l\nu$ control of the systematic 
errors in the quenched approximation has improved. 
Problems remain however, including simulations with pion momentum above 1GeV, 
the chiral extrapolation and unquenching.
An important development is the lattice determination of the zero-recoil form 
factor for $B\rightarrow D^\ast l\nu$. The calculation is systematically 
improvable and with this method it is conceivable that the error on 
${\cal F}_{B\rightarrow D^\ast l\nu}(1)$ can be lowered to under $1\%$.
\section{ACKNOWLEDGEMENTS}
It is a pleasure to thank my collaborators A.\ El-Khadra, A.\ Kronfeld, 
P.\ Mackenzie and J.\ Simone.
I thank
A.\ Ali Khan
S.\ Aoki,
D.\ Becirevic,
C.\ Bernard,
C.\ Davies,
P.\ de Forcrand,
M.\ Garcia-Perez,
J.\ Gill,
S.\ Gottlieb,
S.\ Hashimoto,
Y.\ Kuramashi,
M.\ Kurth,
G.\ Lacagnina,
L.\ Lellouch,
R.\ Lewis,
D.\ Lin,
T.\ Manke,
N.\ Mathur,
C.\ Maynard,
T.\ Naoto,
S.\ Necco,
T.\ Onogi,
J.\ Shigemitsu,
R.\ Sommer,
A.\ Ukawa 
and N.\ Yamada
for discussion and private communication. 

\end{document}